\documentclass[a4paper]{article}
\usepackage{epsfig,amsmath,amsfonts,amssymb,setspace,multirow,textcomp}
\usepackage[T1]{fontenc}
\usepackage[sort&compress]{natbib}
\makeatletter
\renewcommand{\@oddhead}{\textit{Advances in Astronomy and Space Physics} \hfil}
\renewcommand{\@evenfoot}{\hfil \thepage \hfil}
\renewcommand{\@oddfoot}{\hfil \thepage \hfil}
\makeatother

\renewenvironment{thebibliography}[1]{\begin{oldthebibliography}{#1}\setlength{\parskip}{0ex}\setlength{\itemsep}{0ex}}{\end{oldthebibliography}}

\begin{document}
\fontsize{11}{11}\selectfont 
\title{A reconstruction method of electron density distribution in the equatorial region of magnetosphere}
\author{\textsl{V.\,V.~Shastun$^{1}$, O.\,V.~Agapitov$^{2,3}$}}
\date{\vspace*{-6ex}}
\maketitle
\begin{center} {\small $^{1}$LPC2E/CNRS, University of Orleans, France\\
$^{2}$Space Science Laboratory, the University of California 7 Gauss Way, Berkeley, CA 94720, USA\\
$^{3}$Taras Shevchenko National University of Kyiv, Glushkova ave., 4, 03127, Kyiv, Ukraine\\
{\tt vitalii.shastun@cnrs-orleans.fr}}
\end{center}
	
\begin{abstract}
A method for the estimation of electron density from the ratio of the wave magnetic and electric field amplitude of whistler waves is developed. Near the geomagnetic equator, whistler wave normals are mainly close to the direction of the background magnetic field. Dispersion relation  of whistler wave in the parallel propagation approximation is used in this method. Signals registered by STAFF-SA instrument on board the Cluster spacecraft are used for electron density reconstruction. The Cluster spacecraft crossed the plasmasphere at all local times and in a wide range of latitudes over 10 years (2001-2010) covering well the frequency range of both plasmaspheric hiss and lower band chorus emissions in a vicinity of the geomagnetic equator. The proposed technique can be useful in allowing to supplement plasma density statistics obtained from recent probes (such as THEMIS or Van Allen Probes), as well as for reanalysis of  statistics derived from continuous measurements of only one or two components of the wave magnetic and electric fields on board spacecraft covering equatorial regions of the magnetosphere.
\end{abstract}

\section*{\sc Introduction}
\indent \indent The efforts of developing the empirical models that statistically describe the plasmasphere morphology date back to the 1960s, when \citet{binsack1967plasmapause} established an empirical relation of the plasmapause location with the $K_p$ index by using in situ observations from the IMP 2 satellite.
Later on, simultaneous whistler observations from the ground and from the ISEE-1 spacecraft as well as in situ measurements from DE1 and CRRES satellites were used to develop more sophisticated empirical plasmapause models \citep{carpenter1992isee, moldwin2002new, o2003empirical}.
In parallel to the modelling of the plasmapause location, empirical models of the equatorial density have also been developed with whistler observations, in situ probing, radio wave remote measurements, and extreme ultraviolet (EUV) imaging \citep{carpenter1992isee, sheeley2001empirical, denton2004electron, reinisch2004plasmaspheric, song2005magnetospheric, tu2006empirical, larsen2007correlation}.
Moreover, efforts have been made to statistically represent the plasmasphere/plasmatrough density distributions off the geomagnetic equator making use of the measurements from the above-mentioned spacecraft \citep{gallagher1988empirical, gallagher1998simple, gallagher2000global, goldstein2001latitudinal, denton2004electron, huang2004developing, reinisch2004plasmaspheric, tu2006empirical}.

The result of efforts over decades are several commonly used empirical models \citep{sheeley2001empirical, ozhogin2012field}, but our ability to accurately represent the plasmasphere density distribution is still limited. The main reason is that the previous observations have been made in situ along the satellite orbits, and in limited locations under various geomagnetic activities. Existing models are the result of an averaging of the satellite measurements made under different levels of geomagnetic activity. But the plasmasphere is a highly dynamic region that contains density structures of various spatial scales. Its density distributions and their time variations determine the characteristics of the plasma waves (e.g., hiss, chorus, EMIC) within the plasmasphere. Therefore, continuous plasma density measurements for current geomagnetic activity made at high rate are needed.

Here, we present a technique which allows to estimate the electron density $\rho$ from the ratio of the wave magnetic and electric field amplitude of a whistler mode $c|B|/|E|$ in quasi-parallel propagation approximation for cold plasma. Since many probes are equipped for measuring electric and magnetic field amplitudes like multi-component spectral (Cluster STAFF-SA, TC1 STAFF-SA, DE1 PWI, THEMIS FBK) or waveform (POLAR PWI and THEMIS SCM and EFI) measurements and the frequency limitation of multi-component measurements (Akebono PFZ ~400 Hz \citep{kimura1990vlf}, DE1 LFC/PWI \citep{shawhan1981plasma}), our method can be a valuable source of data for cross-calibrating different density measuring equipment and for comparing density profiles obtained by other methods.  This method further can be applied for investigating the wave-normal angle distribution as well as the wave amplitude distribution of the plasmaspheric hiss as functions of magnetic latitude $\lambda$, $L$-shell, Magnetic Local Time (MLT) and geomagnetic activity over a wide region inside the plasmasphere, within $45^\circ$ of the geomagnetic equator.

Our method uses the ratio of magnetic field over electric field components of the hiss and chorus waves. The behaviour of the plasmaspheric hiss and chorus was studied in \citep{hayakawa1986direction, agapitov2011statistical, agapitov2012correction, agapitov2013statistics}. Various statistical studies of the amplitude distribution of hiss waves were presented by \citet{meredith2004substorm} on the basis of CRRES measurements, by \citet{green2005origin} from DE1 measurements, by \citet{kimura1990vlf} from Akebono (Exos-D) data. The amplitude distribution of the hiss waves in the inner belt and slot region as well as accurate assessment of its role in the dynamics of the inner radiation belt have been performed by \citet{agapitov2014inner} using EXOS-D (Akebono) satellite. They all showed that hiss waves are mainly seen from $11:00$ to $19:00$ MLT at $L$-shells from $2.5$ to $3.5$ with a maximum of averaged amplitude of about 20 pT.

The method of the plasmasphere density recovering is based on the field aligned hiss and chorus equatorial distribution approximation. A statistical analysis of the wave normal angle of plasmaspheric hiss was achieved by use of the wave measurements aboard the Akebono spacecraft has been presented by \citet{goto2003determination}. Analysis of the wave distribution function revealed that the distribution of wave normal angles near the magnetic equator depends on $L$-shell: wave normal angles are widely distributed at $L>2.2$, while quasi-parallel propagation is hardly found for $1.2<L<1.8$ \citep{goto2003determination}. A comprehensive statistical study based on Cluster STAFF-SA measurements \citep{agapitov2013statistics} at $L > 2$ showed that the distribution of the angle $\theta$ between the wave normal and background magnetic field at the geomagnetic equator is concentrated in a 30$^{\circ}$ cone, with a maximum around $10^{\circ}-15^{\circ}$ in a vicinity of the geomagnetic equator. The probability density functions of the wave amplitudes and wave-normals are usually non-symmetric and have significant non-Gaussian tails. The plasmaspheric hiss showed a very clear dependence of $\theta$ on $\lambda$ for all wave amplitudes: the mean value of $\theta$ increases with $\lambda$ reaching the resonance cone angle at $\lambda \sim 30^{\circ}-40^{\circ}$. \citet{artemyev2015wave} revealed that a significant fraction of the energy corresponds to very oblique waves with 10 times smaller magnetic power than parallel waves. \citet{mourenas2014consequenses} has shown that very oblique whistler waves have generally a much smaller average intensity than quasi-parallel waves. Nevertheless, very oblique waves with $\theta \sim 60^{\circ}$ dominate pitch angle scattering of energetic electrons. \citet{artemyev2014parametric} calculated the lifetimes of electrons trapped in Earth's radiation belts taking into account quasi-linear pitch-angle diffusion by whistler-mode waves. It was found that analytical lifetimes (and pitch-angle diffusion coefficients) are in good agreement with full numerical calculations based on CRRES and Cluster hiss and lightning-generated wave measurements inside the plasmasphere and Cluster lower-band chorus waves measurements in the outer belt for electron energies ranging from 100 keV to 5 MeV. Therefore significant presence of oblique whistler waves in equatorial region of magnetosphere obligate us to assess the applicability of the approximation of parallel wave propagation for reconstructing electron density.

\section*{\sc Data description and processing technique}

\indent \indent For this work, we made use of the dataset of ELF/VLF waves observed by Cluster from 1 January 2001 to 31 December 2010 in magnetic equatorial region (i.e. confined for the $|\lambda|<5^\circ$) of the plasmasphere and out of it  ($1.5\leq L\leq 5$). Cluster was launched on August 2000. Its mission was investigation of small-scale structures in three dimensions in the Earth's plasma environment, such as those involved in the interaction between the solar wind and the magnetospheric plasma, in global magnetotail dynamics and in cross-tail currents. The four Cluster II spacecraft forms a tetrahedral configuration orbiting into a highly elliptical polar orbit with initial apogee of 19.5 $R_E$ and 4 $R_E$ perigee altitude with a period of 57 hours, permitting measurements at high and low altitudes. Such an elliptical orbit allowed to conduct measurements extending from the magnetotail current sheet, the plasmasphere to the magnetopause.
\begin{figure}
    \centering
    \includegraphics[width=0.75\textwidth]{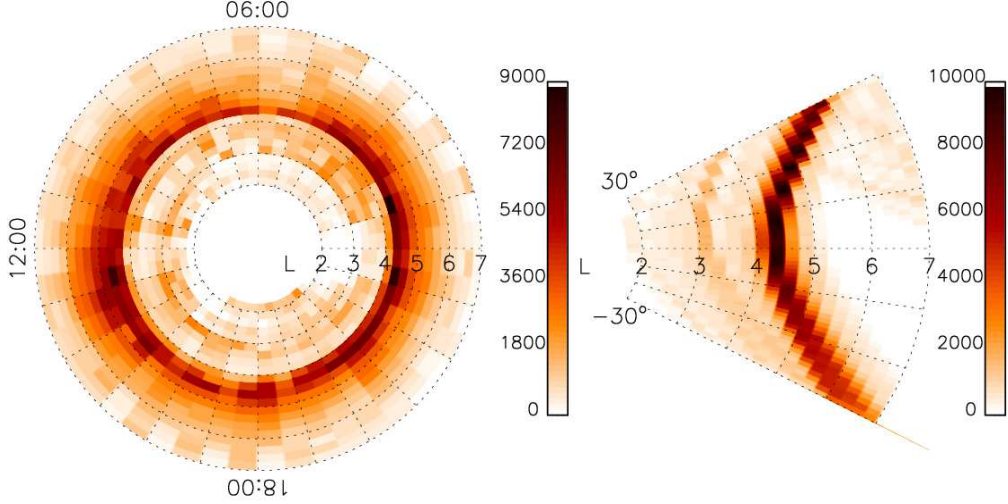}
    \caption{Data coverage of the Cluster STAFF-SA measurements during 2001-2010 in dependence on $L$-shell, $\lambda$ and MLT}
    \label{fig:coverage}
\end{figure}

Our analysis is based on the data from the Spatio-Temporal Analysis of Field Fluctuations - Spectrum Analyzer (STAFF-SA) \citep{cornilleau2003first} and Fluxgate Magnetometer (FGM) \citep{balogh2001cluster,  khotyaintsev2010caabook}. The FGM measures background magnetic field and its low frequency variations. FGM is capable to detect variations of the magnetic field with amplitudes from 0.01 nT to 25000 nT with 4 s time resolution. Noise density is less than 3 pT/Hz$\vphantom{Hz}^{1/2}$. The STAFF-SA instrument has 27 frequency channels with logarithmically spaced central frequencies between 8.8 Hz and 3.56 kHz (covering frequency range from 8 Hz to 4 kHz). STAFF-SA provides the complete spectral matrix (the real and imaginary part) of three magnetic components measured by STAFF search coil magnetometer and two electric components measured by the Electric Fields and Waves instrument  \citep{gustafsson2001first} with a time resolution that is 4 s. The sensitivity of the STAFF search coil magnetometers is $3 * 10^{-3}$ nT Hz$\vphantom{Hz}^{-1/2}$ at 1 Hz, and about $3 * 10^{-5}$ nT Hz$\vphantom{Hz}^{-1/2}$ between 100 Hz and 4 kHz.  Cluster has crossed the plasmasphere at all local times and nearly all latitudes. The Cluster dataset contains a sufficient number of measurement points for performing a statistical study for the considered range of magnetic local times (MLT) and $L$-shells, as illustrated in Fig.~\ref{fig:coverage}. The coverage is very good with only relatively poorer measurements near equator at $L > 5$ and $-10^{\circ} < \lambda < 10^{\circ}$

\citet{agapitov2013statistics} have proposed to check the direct evaluation of the $\theta$-distribution from the ratio of wave magnetic and electric fields $c|B_w|/|E_w|$ measured by STAFF-SA. For each frequency, it can be then estimated from Faraday's Law by use of
\begin{equation}
B_w^2=E_w^2\frac{n^2}{c^2}\sin^2\beta \label{eq:faraday_law}
\end{equation}
where $\beta$ is the angle between $\vec{E}_w$ and  $\vec{k}$ determined from the dispersion matrix and Maxwell equations as:

\begin{equation}
\cos \beta = \frac{\sin \theta
(n^2-P)}{\sqrt{(n^2\sin^2\theta-P)^2+\frac{D^2(n^2\sin^2\theta-P)^2}{(n^2-S)^2}+n^4\sin^2\theta
\cos^2\theta}}, \label{eq:beta_dispersion}
\end{equation}
with $\theta$ -- the angle between the wave normal and background magnetic field.
The refraction index $n$ is estimated from the cold-plasma Appleton-Hartree whistler-mode dispersion relation:

\begin{eqnarray}
 n^2 &=& \frac{{RL\sin ^2 \theta  + PS(1 + \cos ^2 \theta )}}{{2\left( {S\sin ^2 \theta  + P\cos ^2 \theta } \right)}}     \nonumber \\
   &-& \frac{{\sqrt {\left( {RL\sin ^2 \theta  + PS(1 + \cos ^2 \theta )} \right)^2  - 4PRL\left( {S\sin ^2 \theta  + P\cos ^2 \theta } \right)} }}{{2\left( {S\sin ^2 \theta  + P\cos ^2 \theta } \right)}},
\label{eq:refraction_index}
\end{eqnarray}

where $R,L,P$, and $S$ are the Stix parameters \citep{stix1962theory}:

\[
R = 1 -\left( \frac{f_{pe}}{f_{ce}} \right)^2 \left(
\frac{f_{ce}}{f} \right)\left[\left( \frac{f_{ce}}{f}
\right)-1\right]^{-1},
\]

\[
  L = 1 - \left( \frac{f_{pe}}{f_{ce}} \right)^2 \left( \frac{f_{ce}}{f} \right)\left[\left( \frac{f_{ce}}{f} \right)+1\right]^{-1},
\]

and $  S = (R+L)/2,   D = (R-L)/2,   P = 1 - \left(\frac{f_{pe}}{f}\right)^2$, $f_{pe}$ is the local plasma
frequency.

In the approximation of parallel wave propagation $\theta = 0$ the corresponding expression for refraction index $n$ (\ref{eq:refraction_index}) takes form

\begin{equation}
 n^2 = 2S - \sqrt {S^2  - RL}
\label{eq:refractive_index_zero}
\end{equation}

and equation (\ref{eq:beta_dispersion}) simplifies to $\cos\beta = 0$.

\begin{figure}
\centering
\begin{minipage}{.48\textwidth}
  \centering
  \includegraphics[width=1.\linewidth]{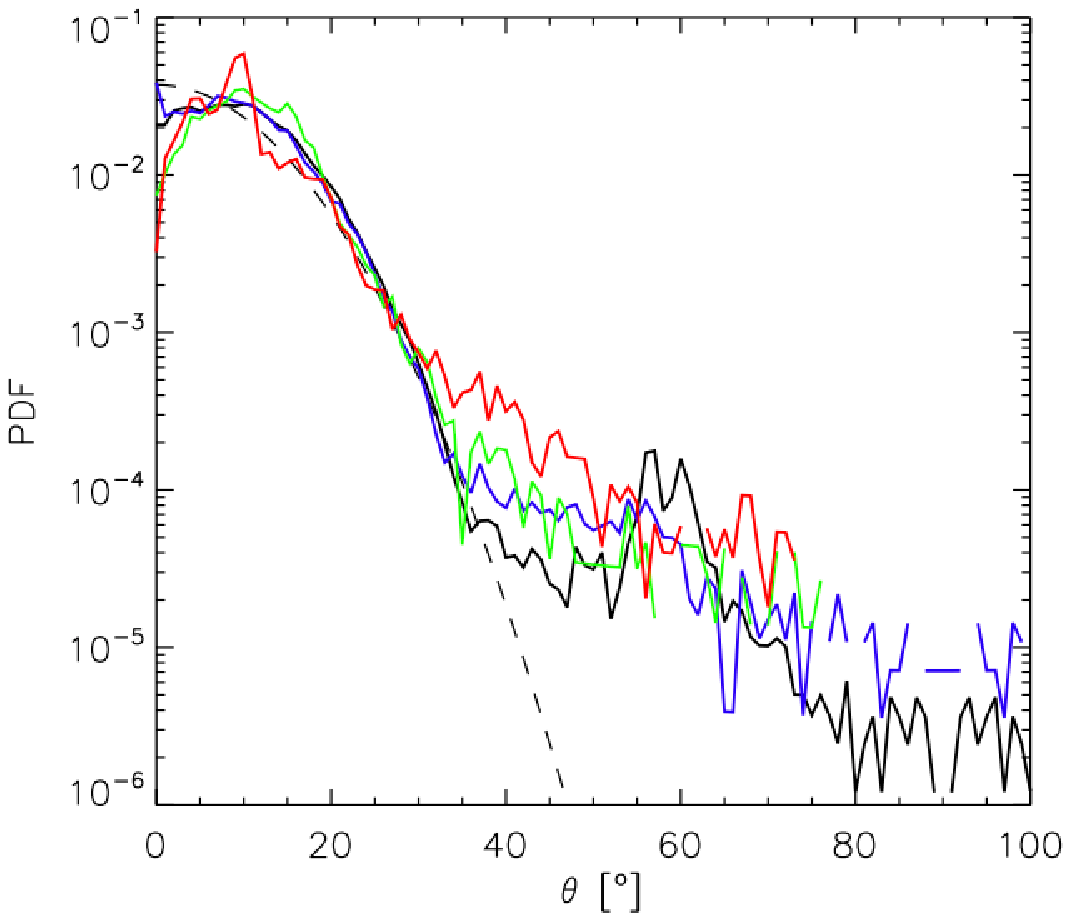}
  \caption{Equatorial distribution of $\theta$ angle for plasmaspheric hiss $(| \lambda | < 5^{\circ})$. Lines indicated by black, green, blue and red colours correspond to $[3,4]$, $[4,5]$, $[5,6]$, $[6,7]$ $L$-shell intervals respectively. The dashed line is the approximation by Gaussian with $\sigma = 14.5^{\circ}$.}
  \label{fig:angle_distribution}
\end{minipage}
\hfill
\begin{minipage}{.48\textwidth}
  \centering
  \includegraphics[width=1.\linewidth]{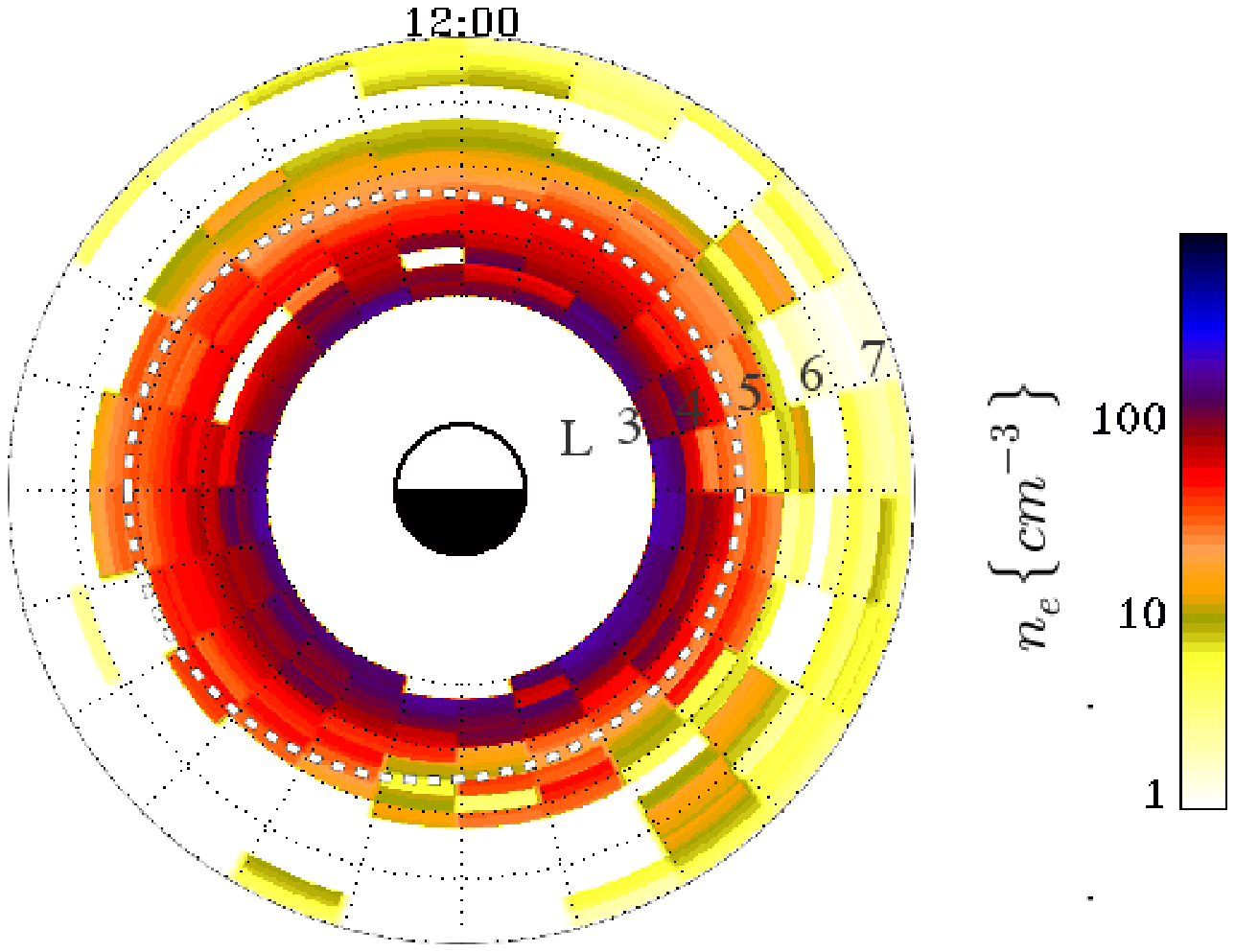}
  \caption{Electron density distribution restored using events during low geomagnetic activity. Modelled plasmapause position is shown by white dashed line.}
  \label{fig3}
\end{minipage}
\end{figure}

After substituting (\ref{eq:refractive_index_zero}) in (\ref{eq:faraday_law}) we obtain equation that can be solved for electron density $\rho$ when $|B_0|$ and $c|B_w| / |E_w|$ are given.

In order to check applicability of the parallel wave approximation we performed statistical analysis of wave normal distribution in equatorial region of magnetosphere for wide range of $L$-shell $(3 < L < 7)$. For data analysis, we used a technique of wave normal vector $\vec{k}$ evaluation under the assumption of single planar wave propagation, as suggested by \citet{means1972use}. It involves the computation of a spectral matrix that consists of the power and the cross-power spectra using three magnetic components. At high signal-to-noise ratios, it was observed that the wave normal direction is well recovered with errors always smaller than $10^{\circ}$ for signal-to-noise ratios greater than one. For the observed chorus waves, this ratio was mostly greater than 1.5. Here we present results obtained by the Means method but verified by the MVAB technique in a way proposed by \citet{agapitov2013statistics}. Fig.~\ref{fig:angle_distribution} shows that wave normals are directed approximately along the background magnetic field in the vicinity of the geomagnetic equator. Mean value of $\theta$ is about $10^{\circ}-15^{\circ}$ and variance $\sigma = 14.5^{\circ}$. The variance of the electron density estimation (the value obtained making use of the parallel wave propagation) caused by the distribution of $\theta$ near the parallel direction (fitted well by the Gaussian with  $\sigma\sim14.5^\circ$, see the dashed curve in Fig.~\ref{fig:angle_distribution}) doesn't exceed $4\%$ of the result. Thus, the $95\%$ confidence interval for electron density estimation with the parallel wave approximation is $8\%$.

Using (\ref{eq:refractive_index_zero}) we present results of the electron density reconstruction in the equatorial region on Fig.~\ref{fig3}. Presented electron density distribution was obtained using the wave magnetic and electric fields $B_w$, $E_w$ obtained from STAFF-SA measurements. Calculation was performed using the data from channels with central frequencies within chorus interval ($0.1f_{ce} < f < f_{ce}$). Background magnetic field $|B_0|$ and local value of gyrofrequency $f_{ce}$ was calculated using the FGM measurements. White dashed line marks the plasmapause location according to global core plasma model during quiet magnetosphere activity developed by \citet{gallagher2000global}. Restored values of the electron density for the plasmasphere region are in good qualitative agreement with widely used empirical models \citep{ozhogin2012field} and \citep{sheeley2001empirical}. The plasmapause is clearly visible in the restored electron density distribution. Its location varies from $L = 4$ in dawn sector to $L = 5$ in dusk sector. Difference of the plasmapause location can be caused by trough in the dusk sector. Overall restored electron density in the plasmasphere region and outer magnetosphere as well as location of the plasmapause correspond to plasma density configuration given by other commonly used empirical models.

\section*{\sc Conclusions}
\indent \indent The technique to reconstruct the electron concentration based on measurements of wave magnetic and electric fields in VLF range and solving the whistler waves dispersion relation is proposed. The equatorial distribution of the electron density has been studied using STAFF-SA VLF measurements on board Cluster spacecraft during 2001--2010. The statistical database spans $L$-shells from 3 to 7 for quiet and active geomagnetic conditions. Comparison of the obtained results with widely used empirical models proposed in \citep{ozhogin2012field} and \citep{sheeley2001empirical} shows that the presented method is consistent with both of these models.

The proposed technique can be useful in allowing to supplement plasma density statistics obtained from recent probes (such as THEMIS or Van Allen Probes), and allows historical reanalysis of electron density derived from continuous measurements of one or two components of the wave magnetic and electric fields on board spacecraft covering equatorial regions of the magnetosphere (i.e. DE-1, THEMIS, POLAR).

\section*{\sc acknowledgement}

\indent \indent The work by V.S. was performed under PROGRESS project funded from the European Union's Horizon 2020 research and innovation programme under grant agreement No 637302. The work by O.A. was performed under JHU/APL Contract No. 922613 (RBSP-EFW).

\end{document}